\documentclass[aps,onecolumn,floatfix,altaffilletter,preprintnumbers,tightenlines,showpacs,showkeys,notitlepage,nofootinbib]{revtex4-2}
\usepackage{amsmath}
\usepackage{amssymb}
\usepackage{titlesec}
\usepackage[a4paper,width=165mm,top=30mm,bottom=30mm]{geometry}
\usepackage{slashed}
\usepackage{braket}
\usepackage{bbm}
\usepackage{cleveref}
\crefname{figure}{Fig.}{Figs.}
\crefname{equation}{Eq.}{Eqs.}
\crefname{section}{Section}{Sections}
\usepackage{xcolor}
\usepackage{graphicx}
\usepackage[caption=false,labelformat=simple]{subfig}
\usepackage{siunitx}
\usepackage{feynmp-auto}

\begin{document}

\title{The Neutrino Mass Bound from Leptogenesis Revisited}

\author{Björn Garbrecht}
\email{garbrecht@tum.de}
\affiliation{Physik-Department T70, Technische Universität München, James-Franck-Straße, 85748 Garching, Germany}

\author{Edward Wang}
\email{edward.wang@tum.de}
\affiliation{Physik-Department T70, Technische Universität München, James-Franck-Straße, 85748 Garching, Germany}

\preprint{TUM-HEP-1535/24}

\begin{abstract}
\noindent
Recent years have seen a great improvement in the computation of $CP$-conserving and $CP$-violating equilibration rates for leptogenesis. These are relevant for the relativistic regime of the sterile Majorana fermions and the dynamics of the Standard Model particles acting as spectator processes. In order to probe the regime of large (${\cal O}(10^2)$) washout parameters, we add $\Delta L = 2$ washout processes, which we derive in the CTP-formalism. To demonstrate their significance, we apply state-of-the-art computational techniques to a simple yet well-motivated phenomenological scenario: unflavored leptogenesis in a hierarchical type-I seesaw model. We then perform a parameter scan of the final baryon asymmetry and find a constraint $m_\text{lightest} \lesssim \SI{0.15}{eV}$ on the absolute neutrino mass scale, which is slightly less stringent than previously reported bounds obtained without the aforementioned improvements. The relaxation of the bounds is mainly due to partially equilibrated spectator fields, which protect part of the asymmetry from washout and lead to larger final asymmetries. While this might seem like a minor correction, the actual dynamics of the fields is substantially altered by these effects. Even though we focused on a particularly simple scenario for leptogenesis, the methods employed here can and should be extended to other models, thus giving us a more accurate picture of the different leptogenesis scenarios.
\end{abstract}

\maketitle

\section{Introduction}

Leptogenesis is a framework that connects two of the long-standing problems of the Standard Model: the origin of neutrino masses and of the baryon asymmetry of the Universe (BAU). If neutrino masses are produced through the coupling to a Majorana fermion, its out-of-equilibrium decay could produce a lepton asymmetry in the early Universe, which would then be converted into a baryon asymmetry via sphaleron processes. One of the first and most compelling proposals to explain the neutrino masses is the seesaw mechanism, in which active neutrinos couple to heavy Majorana fermions via the Higgs boson. One can then find that large Majorana masses naturally explain the smallness of neutrino masses.

The simplest scenario of leptogenesis in the seesaw model is the case of strongly hierarchical Majorana fermions $M_2, M_3 \gg M_1$ without flavour effects. In this setup, an upper bound on the neutrino masses, parametrized by the lightest neutrino mass $m_\text{lightest} \lesssim \SI{0.12}{eV}$ was found~\cite{Buchmuller:2002jk,Buchmuller:2003gz,Giudice:2003jh}. This bound is in agreement with cosmological bounds on neutrino masses, with the constraint $\sum m_\nu < \SI{0.12}{eV}$ (95\% C.L.) from Planck~\cite{Planck:2018vyg}, corresponding to $m_\text{lightest} < 0.03 (0.016) \, \SI{}{eV} $ in normal (inverted) hierarchy, while DESI \cite{DESI:2024mwx} further tightened this constraint to $\sum m_\nu < \SI{0.072}{eV}$ (95\% C.L.), corresponding to $m_\text{lightest} <  \SI{0.0086}{eV} $ in normal hierarchy and below the threshold for inverted hierarchy. However, given the many tensions in cosmological data and between cosmological and terrestrial constraints, the robustness of these bounds is yet to be confirmed~\cite{DiValentino:2021imh,Gariazzo:2023joe,Sekiguchi:2020igz,Forconi:2023akg,Jiang:2024viw}. In view of this, the best model-independent constraint is given by the KATRIN experiment, which placed an upper bound on the effective electron antineutrino mass $m_\text{lightest} \approx m_e = \sqrt{\sum_i |U_{ei}|^2 m_i^2} < \SI{0.8}{eV}$ (90\% C.L.)~\cite{KATRIN:2021uub}. Additionally, in the absence of cancellations due to new physics effects, KamLAND-Zen also places a constraint $m_\text{lightest} < 0.18 - \SI{0.48}{eV}$ (90\% C.L.) from neutrinoless double beta decay assuming Majorana masses~\cite{KamLAND-Zen:2016pfg}. While KamLAND-Zen has since obtained a stronger contraint on the effective Majorana mass $\braket{m_{\beta \beta}}$~\cite{KamLAND-Zen:2022tow}, its translation into a bound on $m_\text{lightest}$ depends on the mass hierarchy. Given recent improvements on the computation of the fluid equations for leptogenesis~\cite{Garbrecht:2019zaa}, it is interesting to investigate  how these affect the predictions of leptogenesis for different neutrino mass scales.

As far as the dynamics of leptogenesis is concerned, a value of $m_\text{lightest} \approx \SI{0.14}{eV}$ that we find in our analysis pushes $M_1 \gtrsim \SI{5e12}{eV}$. At the corresponding temperatures, tau-Yukawa couplings are out of
equilibrium so that leptogenesis is in the unflavoured regime. In view of our interest in the upper bound on $m_\text{lightest}$, for definiteness and for the sake of comparison with earlier papers, we therefore do not include flavour effects in the present analysis, even though they will become relevant on the lower side of the values of $M_1$ that appear allowed by the unflavoured calculation.

The improvements in Ref.~\cite{Garbrecht:2019zaa} mainly focus on two aspects: the careful computation of the rates in the relativistic regime of the lightest Majorana fermion and the inclusion of spectator effects. In the relativistic regime, there are thermal contributions to the rates and to the decay asymmetry that become relevant, and, in addition to this, the interactions are sensitive to the helicity of the Majorana fermions. It is therefore necessary to track the evolution of the different helicity states separately. In the weak washout regime, a significant fraction of this early asymmetry survives at late times, while in the strong washout regime part of the asymmetry gets protected by spectator fields. The precise computation of these early processes is therefore of great relevance for an accurate estimate of the lepton asymmetry at late times.

In Ref.~\cite{Garbrecht:2019zaa}, these methods were applied to a simplified version of the seesaw model to demonstrate the impact of these new effects. In the present work we will apply them to a seesaw model relevant for neutrino mass generation, and investigate the interplay between the parameters for neutrino masses and for leptogenesis. In addition to this, we include a treatment of $\Delta L=2$ processes within the CTP-framework, which was absent in previous works that have introduced an improved treatment of Majorana fermions in the relativistic regimes as well as the dynamics of spectator effects. Subsequent to the analysis of neutrino mass bounds in unflavoured leptogenesis~\cite{Buchmuller:2002jk,Buchmuller:2003gz,Giudice:2003jh}, it has been shown that the aforementioned flavour effects allow one to lower the scale of leptogenesis and to relax many constraints (for a review, see Ref.~\cite{Blanchet:2012bk}). Further, if the reheat temperature is high enough, also the lepton asymmetry from the out-of-equilibrium dynamics of the next-to-lightest sterile Majorana fermion may survive washout from the lightest one and thus directly contribute to the outcome of leptogenesis~\cite{DiBari:2005st}. Yet, the simple unflavoured scenario remains a viable possibility within the seesaw parameter space, and it is illustrative of the relevance of these new methods for phenomenological models of leptogenesis. The applicability of these methods is, however, not restricted to this simplified case, but rather general.

The outline of the article is as follows: in \cref{sec:model} we present the realization of the seesaw model we will employ, and discuss properties of the neutrino mass mechanism as well as of the decay asymmetry of $N_1$ in vacuum within this model. In \cref{sec:fluid} we discuss the fluid equations with the effects described above. In \cref{sec:l=2} we present our treatment of the $\Delta L=2$ processes and in \cref{sec:scan} we show the results from our numerical scan of the asymmetry for different choices of the parameters. We conclude with \cref{sec:conclusions}.

\section{The Model}
\label{sec:model}

The model we consider is the usual type-I seesaw model with three sterile Majorana neutrinos, given by
\begin{equation}
	\mathcal{L} = \mathcal{L}_{SM} + \frac{1}{2} \bar{N}_i i \slashed{\partial} N_i - \frac{1}{2} M_i \bar{N}_i N_i - h_{ij} \bar{\ell}_i \tilde{\phi} N_j - h_{ij}^* \bar{N}_j \tilde{\phi}^* \ell_i,
\end{equation}
where $\tilde{\phi} = i \sigma_2 \phi^*$ is the Higgs doublet conjugated with respect to weak hypercharge and isospin. We further assume strongly hierarchical masses $M_1 \ll M_2 \ll M_3$. After electroweak symmetry breaking, the Higgs field acquires a vacuum expectation value $\sqrt{2} \braket{\phi_0} = v = \SI{246}{GeV}$, producing the Dirac neutrino mass matrix $m_D = h v/\sqrt{2}$. Upon diagonalization of the mass matrix we find the light neutrino mass matrix
\begin{equation}
	m_\nu = m_D M^{-1} m_D^T,
\end{equation}
with real and positive eigenvalues $m_1, m_2$ and $m_3$. In the neutrino mass eigenbasis, one can show that the vacuum $CP$-asymmetry of the $N_1$ decay is given by~\cite{Flanz:1994yx,Covi:1996wh,Buchmuller:1997yu}
\begin{equation}
	\epsilon_0 = \frac{3}{4 \pi} \frac{M_1}{v^2} \sum_{i \neq 1} \frac{\Delta m_{i1}^2}{m_i} \frac{\text{Im} (h_{i1}^2)}{(h^\dagger h)_{11}},
\end{equation}
where $\Delta m_{i1}^2 = m_i^2 - m_1^2$. As shown in Ref.~\cite{Hambye:2003rt}, with the approximation $m_2 \sim m_1$, we can express the maximal asymmetry as
\begin{equation}
	\epsilon_\text{max} = \underset{y}{\text{max}} \frac{3}{16 \pi} \frac{M_1}{v^2} \frac{m_3^2 - m_1^2}{\tilde{m}_1} \text{sinh} \, 2y \sqrt{1 - \left(\frac{2 \tilde{m}_1 - (m_1 + m_3) \text{cosh} \, 2y}{m_3 - m_1} \right)^2}.
	\label{eq:asymmetry}
\end{equation}

It is useful to introduce the washout parameter~\cite{Fry:1980bd,Fukugita:1986hr}
\begin{equation}
	K = \frac{\Gamma_D (z = \infty)}{H (z=1)},
\end{equation}
where $\Gamma_D (z = \infty) = (h^\dagger h)_{11} M_1/(8 \pi)$ is the decay width of $N_1$, as well as the effective neutrino mass~\cite{Plumacher:1996kc}
\begin{equation}
	\tilde{m}_1 = \frac{(m_D^\dagger m_D)_{11}}{M_1},
\end{equation}
which are related as
\begin{equation}
	K = \frac{\tilde{m}_1}{m_*},
\end{equation}
where
\begin{equation}
	m_* = \frac{16 \pi v^2}{m_{Pl}} \sqrt{\frac{g_\star \pi^3}{45}},
\end{equation}
is the equilibrium neutrino mass~\cite{Buchmuller:2004nz,Barbieri:1999ma}, with $m_{Pl} = 1.22 \times \SI{e19}{GeV}$ the Planck mass and $g_\star = 106.75$ the number of relativistic degrees of freedom. The structure of the neutrino mass matrix yields the constraint $\tilde{m}_1 \geq m_\text{lightest}$.

Following the approach from Ref.~\cite{Beneke:2010wd}, we describe the early Universe by a spatially flat Friedman-Lemaître-Robertson-Walker (FLRW) metric. We can then shift our description into comoving coordinates, with comoving quantities defined as $\vec{k} = a(t) \vec{k}_\text{phys}$, $T = a(t) T_\text{phys}$ and $s = a(t)^3 s_\text{phys}$, where $a(t)$ is the scale factor of the FLRW metric. Here, $T$ is the temperature, $\vec k$ momentum and $s$ the entropy densisty; and we have indicated the physical parameters with a subscript `phys'. Introducing the conformal time $\eta$, related to $t$ by $dt = a d \eta$, the scale factor in the radiation-dominated Universe is given by $a = a_R \eta$. We choose
\begin{equation}
	a_R = T = \frac{m_{Pl}}{2} \sqrt{\frac{45}{g_\star \pi^3}}.
\end{equation}
In this way, the effect of the expansion of the Universe is described by masses that scale with $a$. We further introduce the dimensionless time variable $z = M_1/T_\text{phys}$. With this we can write the washout parameter as
\begin{equation}
	K = \frac{g_w a_R (h^\dagger h)_{11}}{16 \pi M_1}.
\end{equation}
One commonly distinguishes the scenarios of strong and weak washout, characterized by the conditions $K > 1$ and $K < 1$ respectively.

Additionally, we approximate the Standard Model particles as being in kinetic equilibrium, so that the charge asymmetries can be parametrized by their respective chemical potentials. For small chemical potentials and relativistic particles, we have
\begin{equation}
	\mu_X \approx \frac{3 g_s s}{T^2} Y_X,
\end{equation}
with $g_s = 2$ for fermions and $g_s = 1$ for scalars.

\section{Fluid Equations for Leptogenesis}
\label{sec:fluid}

In our description of leptogenesis we employ momentum-averaged fluid equations, which in general leads to an order one uncertainty in the final asymmetry~\cite{Ghiglieri:2017csp,Asaka:2011wq}; however, this approximation works well in the strong washout regime due to the nonrelativistic kinematics of the sterile Majorana fermions~\cite{Basboll:2006yx,Hahn-Woernle:2009jyb}. Throughout this work we will use the Schwinger-Keldysh Closed-Time-Path (CTP) formalism~\cite{Schwinger:1960qe,Keldysh:1964ud} applied to nonequilibrium quantum-field theory~\cite{Calzetta:1986cq}. This allows for a self-consistent tratment of the rates from first principles.

Leptogenesis can be described by a set of coupled kinetic equations, which, in its simplest scenario, is given by
\begin{subequations}
\label{eq:fluid_both}
\begin{align}
	\frac{\text{d}}{\text{d} z} Y_{N 1} &= - \Gamma_D (Y_{N 1} - Y_{N 1}^\text{eq}), \\
	\frac{\text{d}}{\text{d} z} Y_\ell &= - \epsilon_0 \Gamma_D (Y_{N 1} - Y_{N 1}^\text{eq}) - \Gamma_W Y_\ell,
	\label{eq:kinetic_minimal}
\end{align}
\end{subequations}
where we define the particle yields
\begin{equation}
	Y_X = \frac{1}{s} (n_X^+ - n_X^-),
\end{equation}
with $n_X^{+, -}$ the particle and antiparticle number densities for species $X$. For Majorana fermions, for which the distinction between particle and antiparticle does not apply, we instead use
\begin{equation}
	Y_N = \frac{1}{s} n_N.
\end{equation}
In \cref{eq:kinetic_minimal}, $\Gamma_W$ are washout rates, which comprise inverse decays and lepton number violating scatterings.

As discussed in Ref.~\cite{Garbrecht:2019zaa}, the description through Eqs.~(\ref{eq:fluid_both}) are incomplete for a number of reasons. The first is the handling of the Majorana particles. The decay rate $\Gamma_D$ is the vacuum decay rate of Majorana particles at rest. As the authors of that paper showed through direct numerical comparison, this rate is sufficiently accurate for $z \gtrsim 10$. For $z \lesssim 10$, however, one needs to take into account the dilation of the decay rate, and for $z \lesssim 1$ thermal effects also become relevant. The second issue is that Eqs.~(\ref{eq:fluid_both}) make no distinction between the helicity states of the Majorana fermion, which is relevant for relativistic particles. Both of these issues have been addressed in Ref.~\cite{Garbrecht:2019zaa} in the derivation of the rates in the CTP formalism.

We denote by $\ell_\parallel$ the linear combination of leptons that couple to $N_1$, by $B$ baryon and $L$ lepton number. Noting that in the temperature range of interest above $10^{13}{\rm GeV}$ lepton flavour is conserved and assuming the particles $N_1$ are the only sourse of asymmetry, we can set $Y_{\Delta_\parallel} = Y_{B/3} - 2 Y_{\ell_\parallel} = Y_{B-L}$. Further, we define the even and odd combinations of helicity
\begin{equation}
	Y_{N \text{even/odd}} = \frac{1}{s} (n_{N +} \pm n_{N -}).
\end{equation}
In terms of these quantities, one arrives at the relativistic fluid equations
\begin{subequations}
\begin{align}
	\frac{\text{d}}{\text{d} z} Y_{N_1 \text{even}} &= - \Gamma \cdot (Y_{N_1 \text{even}} - Y_{N_1 \text{eq}}), \\
	\frac{\text{d}}{\text{d} z} Y_{N_1 \text{odd}} &= - \Gamma \cdot Y_{N_1 \text{odd}} - \eta_{N_1} \tilde{\Gamma} \cdot (Y_{\ell_\parallel} + \frac{1}{2} Y_\phi), \\
	\frac{\text{d}}{\text{d} z} Y_{\Delta_\parallel} &= \tilde{\Gamma} \cdot Y_{N_1 \text{odd}} - \epsilon_\text{eff} \Gamma \cdot (Y_{N_1 \text{even}} - Y_{N_1 \text{eq}}) + \eta_{N_1} \Gamma \cdot (Y_{\ell_\parallel} + \frac{1}{2} Y_\phi).
\end{align}
\end{subequations}
Here we have introduced the rates
\begin{equation}
	\Gamma = K \frac{1}{2} (\gamma_\text{LNC} + \gamma_\text{LNV}), \quad \tilde{\Gamma} = K \frac{1}{2} (\gamma_\text{LNC} - \gamma_\text{LNV})
\end{equation}
and the effective $CP$-violating parameter at finite temperature
\begin{equation}
	\epsilon_\text{eff} = \epsilon_0 \frac{\mathcal{K} (z)}{\mathcal{I}(z)} \frac{2 \gamma_\text{LNC} \cdot \gamma_\text{LNV}}{z^2 (\gamma_\text{LNC} + \gamma_\text{LNV})},
\end{equation}
which depend on the lepton-number conserving and violating rates $\gamma_\text{LNC}$ and $\gamma_\text{LNV}$ respectively, and with
\begin{align}
	\mathcal{I} (z) &= \int_z^\infty dy \frac{y \sqrt{y^2 - z^2}}{e^y + 1} = z^2 \sum_{n=1}^\infty \frac{(-1)^{n+1}}{n} K_2 (n z) \approx z^2 K_2 (z), \\
	\mathcal{J} (z) &= \int_z^\infty dy \frac{y \sqrt{y^2 - z^2} e^y}{(e^y + 1)^2} = z^2 \sum_{n=1}^\infty (-1)^{n-1} K_2 (n z) \approx z^2 K_2 (z), \\
	\mathcal{K} (z) &= \int_z^\infty dy \frac{y^2 \sqrt{y^2 - z^2}}{e^y + 1} = \sum_{n=1}^\infty (-1)^{n+1} \left(\frac{z^3}{n} K_1 (n z) + \frac{3 z^2}{n^2} K_2 (n z) \right) \approx z^3 K_1 (z) + 3 z^2 K_2 (z).
\end{align}
In terms of these we can also write
\begin{equation}
	Y_{N_1 \text{eq}} (z) = \frac{T^3}{s \pi^2} \mathcal{I} (z), \quad \eta_{N_1} (z) = \frac{6}{\pi^2} \mathcal{J} (z).
\end{equation}
In the case of fully equilibrated spectators, that is strong and weak sphalerons as well as top and bottom-Yuakwa interactions in the temperature range of interest, one obtains the relations
\begin{equation}
	Y_{\ell_\parallel} = - \frac{13}{30} Y_{\Delta_\parallel}, \quad Y_\phi = - \frac{1}{5} Y_{\Delta_\parallel}.
\end{equation}

We take the numerical data points for $\gamma_{\text{LNC}}$ and  $\gamma_{\text{LNV}}$ obtained in Ref.~\cite{Garbrecht:2019zaa} and interpolate between them to compute our rates. Note that for $z\to 0$, $\gamma_{\text{LNV}}$ vanishes because the lepton-number violation through the Majorana mass $M_1$ becomes irrelevant at high temperatures. For $z\gtrsim 1$, one recovers $\gamma_{\text{LNC}}\approx\gamma_{\text{LNV}}$, indicating that a $1\leftrightarrow 2$ process between $N_1$, $\ell$ and $\phi$ violates lepton number at a coin-toss chance. As for the parameter $\epsilon_{\rm eff}$, sizeable early asymmetries can be produced at small $z$ in spite of the suppression of  $\gamma_{\text{LNV}}$ because it is compensated by a large deviation of $N_1$ from equilibrium---an effect more sizeable for vanishing than for thermal initial conditions of $N_1$.

While these relativistic effects are irrelevant if the early asymmetries are destroyed by strong washout, they can have a sizeable effect in the weak washout regime or in the presence of partially equilibrated spectator fields which protect some of the asymmetry from washout. To describe the spectator fields we introduce the quark yields $Y_{Q_i}$ for the left-handed doublets, and $Y_{u_i}, Y_{d_i}, Y_t, Y_b, i=1,2$ for the right-handed singlets, as well as the lepton fields not coupling to $N_1$ directly $Y_{\ell_\perp 1}, Y_{\ell_\perp 2}$. Since at high temperatures the lepton and the first and second quark generation Yukawa couplings are negligible, we can set $Y_{u_i} = Y_{d_i} = Y_d$, $Y_{Q_1} = Y_{Q_2} = Y_Q$ and $Y_{\ell_\perp 1} = Y_{\ell_\perp 2} = Y_{\ell_\perp}$. At $T \sim \SI{e13}{GeV}$, the relevant partially equilibrated interactions are bottom-Yukawa and weak sphaleron interactions. Defining $Y_{\Delta \text{down}} = Y_b - Y_d$, the full system of Boltzmann equations is~\cite{Garbrecht:2014kda}
\begin{subequations}
\begin{align}
	\frac{\text{d}}{\text{d} z} Y_{N_1 \text{even}} &= - \Gamma \cdot (Y_{N_1 \text{even}} - Y_{N_1 \text{eq}}), \\
	\frac{\text{d}}{\text{d} z} Y_{N_1 \text{odd}} &= - \Gamma \cdot Y_{N_1 \text{odd}} - \eta_{N_1} \tilde{\Gamma} \cdot (Y_{\ell_\parallel} + \frac{1}{2} Y_\phi), \\
	\frac{\text{d}}{\text{d} z} Y_{\Delta_\parallel} &= \tilde{\Gamma} \cdot Y_{N_1 \text{odd}} - \epsilon_\text{eff} \Gamma \cdot (Y_{N_1 \text{even}} - Y_{N_1 \text{eq}}) + \eta_{N_1} \Gamma \cdot (Y_{\ell_\parallel} + \frac{1}{2} Y_\phi), \\
	\frac{\text{d}}{\text{d} z} Y_{\Delta \text{down}} &= - \Gamma_\text{down} \cdot (Y_b - Y_{Q_3} + \frac{1}{2} Y_\phi), \label{eq:hbottom}\\
	\frac{\text{d}}{\text{d} z} Y_{\ell_\perp} &= - \Gamma_\text{ws} \cdot (9 Y_{Q_3} + 18 Y_Q + 3 Y_{\ell_\parallel} + 6 Y_{\ell_\perp}),
\end{align}
\end{subequations}
with the equilibration rates~\cite{Garbrecht:2014kda,Garbrecht:2013bia,Moore:2000ara}
\begin{equation}
	\Gamma_\text{down} \approx 1.0 \times 10^{-2} \frac{h_b^2 T}{M_1}, \quad \Gamma_\text{ws} \approx (8.24 \pm 0.10) \left( \text{log} \left(\frac{m_D}{g_2^2 T} \right) + 3.041 \right) \frac{g_2^2 T^3}{2 m_D^2 M_1} \alpha_2^5,
\end{equation}
with $h_b$ the bottom-Yukawa coupling, $\alpha_2 = g_2^2/4 \pi$ the $SU(2)_L$ electroweak coupling strength and $m_D^2 \approx \frac{11}{6} g_2^2 T^2$ the thermal mass of the $SU(2)_L$ gauge bosons.

In addition to this, we need to relate the yields $Y_{\ell_\parallel}, Y_{Q_3}, Y_b, Y_Q, Y_\phi$ to $Y_{\Delta_\parallel}, Y_{\ell_\perp}, Y_{\Delta_\text{down}}$ in order to obtain a closed system of equations. From the constraints on the chemical potentials following from the top-quark Yukawa-couplings and strong sphalerons being in equilibrium, we find the relations
\begin{equation}
\renewcommand*{\arraystretch}{1.2}
\left(\begin{array}{c}
		Y_{l_\parallel}\\
		Y_{Q_3}\\
		Y_b\\
		Y_Q\\
		Y_\phi
\end{array}\right)\  =
	\begin{pmatrix}
-\frac{1}{2} 	& 1 			& 0 				\\
\frac{1}{23} 	& \frac{1}{2} 	& -\frac{10}{23} 	\\
\frac{1}{46} 	& \frac{1}{2} 	& \frac{18}{23} 	\\
-\frac{1}{46} 	& \frac{1}{2} 	& \frac{5}{23} 		\\
-\frac{7}{23} 	& 0 			& \frac{24}{23} 	
	\end{pmatrix} 
\left(\begin{array}{c}
Y_{\Delta_\parallel}\\
Y_{l_\perp}\\
Y_{\Delta_\text{down}}
\end{array}\right)\ .
\end{equation}

In the present scenario, the early asymmetries in $Y_\phi$ are transferred through the $B$ and $L$-conserving bottom-Yukawa couplings to $Y_{\Delta\rm down}$, as described by \cref{eq:hbottom}. There, the asymmetry is effectively hidden from washout through $N_1$, as long as the bottom-Yukawa coupling doesn't fully equilibrate. The down-quark asymmetry maintains an asymmetry in $\phi$, which in turn creates a bias in $\ell_\parallel$ that  is not fully erased as lepton-number violation through $N_1$ freezes out. This way, an early asymmetry in $\phi$, which is neither baryonic nor leptonic, turns into a lepton asymmetry during the freeze-out of $N_1$.

\section{$\Delta L = 2$ Processes}
\label{sec:l=2}

In Ref.~\cite{Garbrecht:2019zaa}, only the on-shell part of the $N_1$ propagator was considered, due to the smallness of the propagator width. In doing so, however, one neglects $\Delta L=2$ contributions to the washout, which limit the lepton asymmetry for large Yukawa couplings and $M_1$ and are therefore paramount for the determination of the bound on $m_{\rm lightest}$. In this section, we derive the $\Delta L=2$ contribution to the washout by adding a purely off-shell part to the $N_1$ spectral self-energy. This off-shell part is due to loop insertions to the propagator, where the particles running in the loop are on-shell, even if the propagator itself is not. When inserting these terms into the collision term, we find that this corresponds to $\Delta L = 2$ scattering processes. In this way, we are able to determine these rates in an entirely consistent way within the CTP formalism.

One crucial difference between our result and the one from Ref.~\cite{Buchmuller:2002rq} is that the authors of that paper considered a washout rate averaged over all lepton flavours. When doing so and adding the interactions with all heavy Majorana fermions, a cancellation of terms similar to the one in the neutrino mass matrix occurs, such that the final rate does not depend on the specific form of the Yukawa matrix or the washout parameter $K$ but only on the values of the neutrino masses. We improve on this approach, taking into account that the $N_1$ decays don't produce all lepton flavours equally, but only a specific linear combination of flavours. The interactions of $\ell_\parallel$ with the heavier Majorana fermions heavily depend on the form of the Yukawa matrix, and so the cancellation they reported doesn't occur in this case. Since we expect these interactions to be subdominant with respect to the interactions with $N_1$ due to their large masses, we only keep the latter ones. Even though we are considering less interaction terms than the previous analysis, for large $K$, the washout rates we find are larger than previously found. The reason for this is that, in the flavour-averaged case, large $K$ terms are always cancelled by the $N_2$ and $N_3$ interaction terms. In our case, we know that this cancellation is not exact in general and so we expect a large washout rate to remain even if we were to include interactions with $N_2$ and $N_3$ in our analysis.

At a methodical level, the leading order evaluation of the $CP$ asymmetry in the CTP formalism readily yields Boltzmann equations that predict no $CP$ asymmetry in thermal equilibrium. We therefore need to include self-energy corrections in the spectral propagator of the Majorana neutrino to capture off-shell effects, as we shall do in the following. For comparison, in conventional setups where S-matrix elements are substituted into Boltzmann equations, the evaluation based on $1\leftrightarrow 2$ processes and their inverse only misses leading order contributions that guarantee $CP$-symmetric  equlibrium conditions. The missing contributions are then attributed to $2\leftrightarrow 2$ processes, where overcounting of reactions mediated by on-shell Majorana neutrinos is dealt with by the subtraction of real intermediate states. This approach has been worked out in Ref.~\cite{Kolb:1979qa} for general scenarios of baryogenesis from out-of-equilibrium decays. It then has been applied to leptogenesis in Ref.~\cite{Buchmuller:1997yu}, and further detailed discussion can be found in Refs.~\cite{Buchmuller:2002rq,Giudice:2003jh,Buchmuller:2004nz}.

Similarly to the case of Standard Model leptons, the $N$ propagator in kinetic equilibrium can be written as
\begin{equation}
\label{f:less:finite:width}
	i S_N^< (k) = - 2 S_N^\mathcal{A} f_N (k), \quad i S_N^> (k) = 2 S_N^\mathcal{A} (1 - f_N (k)).
\end{equation}
The spectral propagator for a massive fermion, when summing up the self energies, is given by
\begin{equation}
S_{N_1}^{\mathcal{A}} (k) = \left[\left(\slashed{k} + M_1 - {\Sigma}^\mathcal{H}_N \left(k \right) \right) \cdot \frac{\Gamma_N}{\Omega_N^2 + \Gamma_N^2}
- {\Sigma}^\mathcal{A}_N\left(k \right) \frac{\Omega_N}{\Omega_N^2 + \Gamma_N^2} \right],
\label{eq:resummed_propagator}
\end{equation}
with 
\begin{equation}
\Gamma_N \left(k\right) = 2 \left( k_{\mu} - \Sigma^\mathcal{H}_{N, \mu} \right) \cdot \Sigma^{\mathcal{A},\mu}_{N}, \quad \Omega_N \left(k\right) = \left( k_{\mu} - \Sigma^\mathcal{H}_{N, \mu} \right)^2  - M_1^2 - \left(\Sigma^{\mathcal{A}}_{N,\mu}\right)^2.
\end{equation}
As a subtlety, note that when $\phi$ and $\ell$ are in equilibrium, then strictly speaking only the Fermi--Dirac part of the distribution $f_N$ in \cref{f:less:finite:width} goes with the finite width form of the spectral function from \cref{eq:resummed_propagator}, while the out-of-equlibrium remainder should go with a sharply peaked on-shell $\delta$-distribution. However, we may still effectively use the form as in \cref{f:less:finite:width} because the $\delta$-distribution is shifted to a finite-width form when all gradients incurred as $N_1$ relaxes toward equilibrium are properly taken into account~\cite{Garbrecht:2011xw}.
The $N_1$ self-energy can be decomposed as
\begin{equation}
	i \Sigma_{N_1}^{>,<} (k) = g_w (h^\dagger h)_{11} (P_L \gamma_\mu i \hat{\Sigma}_{N_1, L}^{\mu >,<} (k) + P_R \gamma_\mu i \hat{\Sigma}_{N_1, R}^{\mu >,<} (k)),
\end{equation}
with
\begin{align}
	i \hat{\Sigma}_{N_1, L}^{\mu >,<} (k) &= \frac{1}{2} \int \frac{d^4 p}{(2 \pi)^4} \text{tr} [\gamma^\mu P_L i S_{\ell_\parallel}^{>,<} (p) P_R] i \Delta_\phi^{>,<} (k-p), \\
	i \hat{\Sigma}_{N_1, R}^{\mu >,<} (k) &= \frac{1}{2} \int \frac{d^4 p}{(2 \pi)^4} \text{tr} [\gamma^\mu P_R C( i S_{\ell_\parallel}^{<,>} (-p))^T C^\dagger P_L] i \Delta_\phi^{<,>} (p-k).
\end{align}
In kinetic equilibrium, they satisfy the generalized Kubo--Martin--Schwinger (KMS) relations
\begin{equation}
	\hat{\Sigma}_{N_1, L/R}^> (k) = - e^{(k_0 \mp \mu_{\ell} \mp \mu_\phi)/T} \hat{\Sigma}_{N_1, L/R}^< (k),
\end{equation}
and we can express the spectral self energies as
\begin{subequations}
\begin{align}
	\hat{\Sigma}_{N_1, L/R}^{\mathcal{A} 0} (k) =& \frac{T^2}{16 \pi |\mathbf{k}|} I_1 \left(\frac{k^0}{T}, \frac{|\mathbf{k}|}{T}, \pm \frac{\mu_\ell}{T}, \pm \frac{\mu_\phi}{T} \right), \\
	\hat{\Sigma}_{N_1, L/R}^{\mathcal{A} i} (k) =& \frac{T^2}{16 \pi |\mathbf{k}|} \frac{k^i}{|\mathbf{k}|} \left[ \frac{k^0}{|\mathbf{k}|}  I_1 \left(\frac{k^0}{T}, \frac{|\mathbf{k}|}{T}, \pm \frac{\mu_\ell}{T}, \pm \frac{\mu_\phi}{T} \right) - \frac{(k^0)^2 - \mathbf{k}^2}{2 |\mathbf{k}| T} I_0 \left( \frac{k^0}{T}, \frac{|\mathbf{k}|}{T}, \pm \frac{\mu_\ell}{T}, \pm \frac{\mu_\phi}{T} \right) \right],
\end{align}
\end{subequations}
with
\begin{subequations}
\begin{align}
	\nonumber  I_0 (y_0, y, u_1, u_2) =& (2 u_1 - y_0) \theta (y^2 - y_0^2) + \text{log} \left(\frac{1-e^{-(|y_0+y|)/2+(-1)^{\theta(-y_0 - y)} u_2}}{1-e^{-(|y_0-y|)/2 + (-1)^{\theta(y-y_0)} u_2}} \right) \\
	&+ \text{log} \left(\frac{1+e^{(|y_0+y|)/2+(-1)^{\theta(y_0 + y)} u_1}}{1+e^{(|y_0-y|)/2+(-1)^{\theta(y_0 - y)} u_1}} \right), \\
	\nonumber I_1 (y_0, y, u_1, u_2) =& \frac{y |y_0|}{2} \theta (y_0^2 - y^2) + \frac{\pi^2 + u_1^2 - u_2^2 - \text{sign} (y_0) (|y_0| - y) (u_1 - u_2)}{2} \theta(-y_0^2 + y^2) \\
	\nonumber & + \frac{y_0+y}{2} \text{log} \left(\frac{1+e^{-(|y_0+y|)/2+(-1)^{\theta(-y_0 - y)} u_1}}{1-e^{-(|y_0-y|)/2+(-1)^{\theta(y-y_0)} u_2}} \right) - \frac{y_0-y}{2} \text{log} \left(\frac{1+e^{-|y_0-y|/2+(-1)^{\theta(y-y_0)} u_1}}{1-e^{-|y_0+y|/2+(-1)^{\theta(-y_0 - y)} u_2}} \right) \\
	\nonumber & + \text{Re} \bigg[\text{Li}_2 \left(-e^{-(y_0-y)/2+\text{sign} (y_0) u_1} \right) - \text{Li}_2 \left(e^{-(y_0-y)/2+\text{sign} (y_0) u_2} \right) \\
	& - \text{Li}_2 \left(-e^{-(y_0+y)/2+\text{sign} (y_0) u_1} \right) + \text{Li}_2 \left(e^{-(y_0+y)/2+\text{sign} (y_0) u_2} \right) \bigg],
\end{align}
\end{subequations}
for massless particles running in the loop. This generalizes the results from Refs.~\cite{Garbrecht:2013bia,Glowna:2015aos,Garbrecht:2013gd} for particles with chemical potentials in the loop and reduces to the previously known results in the case of vanishing chemical potentials.

\begin{figure}[t]
\centering
\subfloat[]{
\centering
\includegraphics[width=0.48\textwidth]{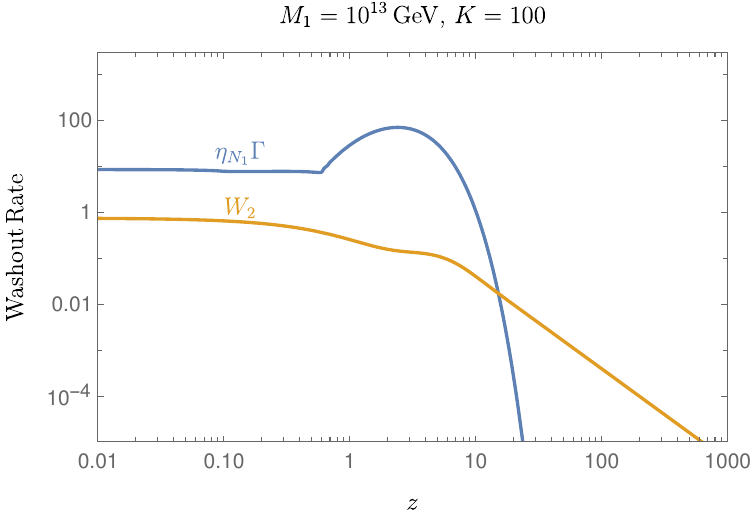}
\label{fig:rates_100}}
\subfloat[]{
\centering
\includegraphics[width=0.48\textwidth]{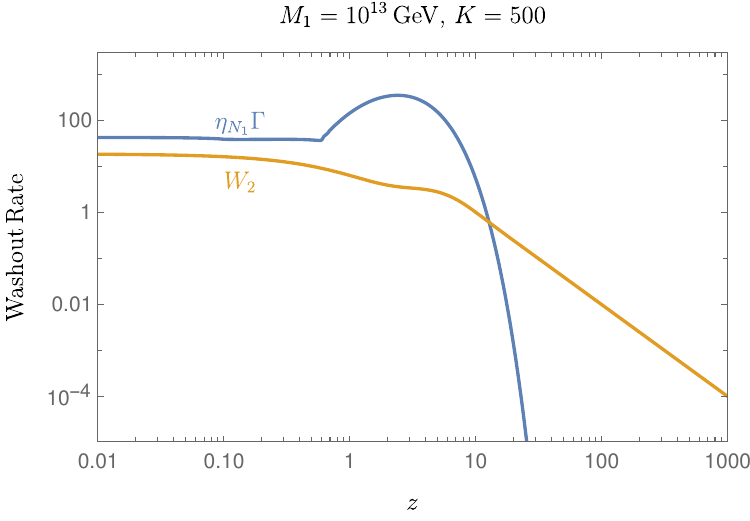}
\label{fig:rates_500}}
\caption{Comparison of washout rates for $\tilde{M}_1 = \SI{e13}{GeV}$ for $K=100$ (a) and $K=500$ (b).}
\label{fig:rates}
\end{figure}

As in Ref.~\cite{Garbrecht:2019zaa}, the collision term for charged leptons without the $CP$-violating source term is given by
\begin{equation}
	g_w \frac{\text{d}}{\text{d} z} Y_{\ell_\parallel} = \frac{1}{\tilde{M}_1 s} \int \frac{d^4 k}{(2 \pi)^4} \text{tr} [i S_{N_1}^> (k) P_L i \Sigma_{N_1}^< (k) - i S_{N_1}^< (k) P_L i \Sigma_{N_1}^> (k)].
	\label{eq:collision_term}
\end{equation}
Since in the relevant parameter region the Yukawa couplings are comparatively small, we can neglect the width of $N_1$ while retaining the off-shell part of the propagator. We can then write\footnote{In principle, also the first term in the propagator contains an off-shell contribution to the collision term. However, this term is more strongly peaked around the pole than the second one, and it does not depend on the chemical potentials to leading order. We therefore neglect this contribution.}
\begin{equation}
	S_{N_1}^{\mathcal{A}} (k) \approx \pi \delta(k^2 - M_1^2) \text{sign} (k^0) (\slashed{k} + M_1) - \frac{\Sigma_N^\mathcal{A} (k)}{k^2 - M_1^2}.
\end{equation}
Taking the off-shell part of the propagator and using the KMS relation for the self-energy, we can write
\begin{equation}
	g_w \frac{\text{d}}{\text{d} z} Y_{\ell_\parallel} =- \frac{8 g_w^2 (h^\dagger h)_{11}^2}{\tilde{M}_1 s} \int \frac{d^4 k}{(2 \pi)^4} (f_F (k_0 - \mu_\ell - \mu_\phi) - f_F (k_0 + \mu_\ell + \mu_\phi)) \frac{1}{k^2 - M_1^2} \hat{\Sigma}_{N_1, R \, \mu}^\mathcal{A} (k) \hat{\Sigma}_{N_1, L}^{\mu \mathcal{A}} (k),
	\label{eq:collision_expanded}
\end{equation}
which, to first order in the chemical potentials, is
\begin{equation}
	g_w \frac{\text{d}}{\text{d} z} Y_{\ell_\parallel} =- \frac{16 g_w^2 (h^\dagger h)_{11}^2}{\tilde{M}_1 s} \frac{\mu_\ell + \mu_\phi}{T} \int \frac{d^4 k}{(2 \pi)^4} (1-f_F (k_0)) f_F (k_0) \frac{1}{k^2 - M_1^2} [\hat{\Sigma}_{N_1, R \, \mu}^\mathcal{A} (k) \hat{\Sigma}_{N_1, L}^{\mu \mathcal{A}} (k)]_{\mu_\ell = \mu_\phi = 0}.
\end{equation}
From this we can extract the new contribution to the washout rate
\begin{equation}
	W_2 = -\frac{24 K^2}{T^3} \frac{\tilde{M}_1}{T} \, \left( Y_{l_{\parallel}} + \frac{1}{2} Y_{\phi} \right) \int \frac{\text{d}^4 k}{(2\pi)^4} ( 1 - f_F(k_0) ) f_F(k_0)\, \gamma_{\Delta L = 2} (k), 
\end{equation}
with
\begin{equation}
	\gamma_{\Delta L=2} (k) = -\frac{(32 \pi)^2}{T} \frac{1}{k^2 - M_1^2} \hat{\Sigma}_{N_1 \mu} (k) \hat{\Sigma}_{N_1}^{\mu}(k),
\end{equation}
which correspond to the $\Delta L = 2$ scattering processes. Note that, since the chemical potentials appear in both $\Sigma_{N_1, L}$ and $\Sigma_{N_1, R}$, when expanding in the chemical potentials we pick up a factor of 2 compared to if we had only considered the potentials from either particles or antiparticles. Since the part of the propagator producing these processes is purely off shell, no real-intermediate-state subtraction is necessary. As expected, these processes do not produce a backreaction into $Y_{N_1 \text{odd}}$ since  $\text{tr} [\text{P}_h \gamma^5 \gamma^\mu \Sigma_{N_1}^\mathcal{A}] \Sigma_{N_1 \mu}^\mathcal{A} = 0$.

\begin{figure}[t]
\centering
\subfloat[]{
\centering
\includegraphics[width=0.48\textwidth]{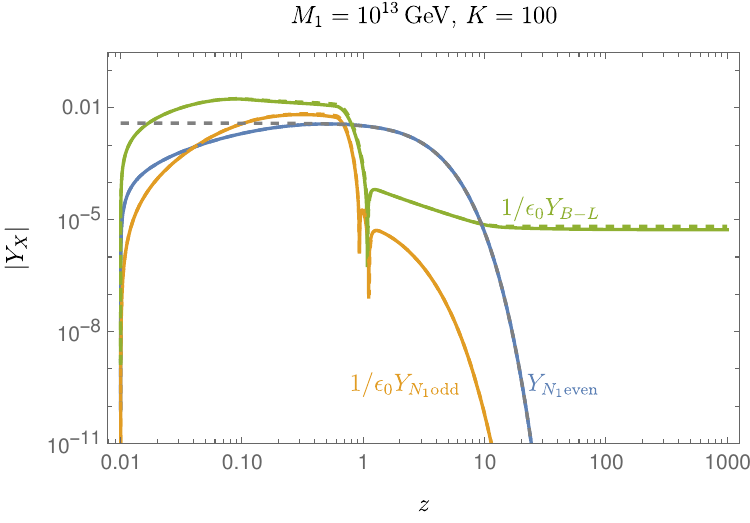}
\label{fig:washout_100}
}
\subfloat[]{
\centering
\includegraphics[width=0.48\textwidth]{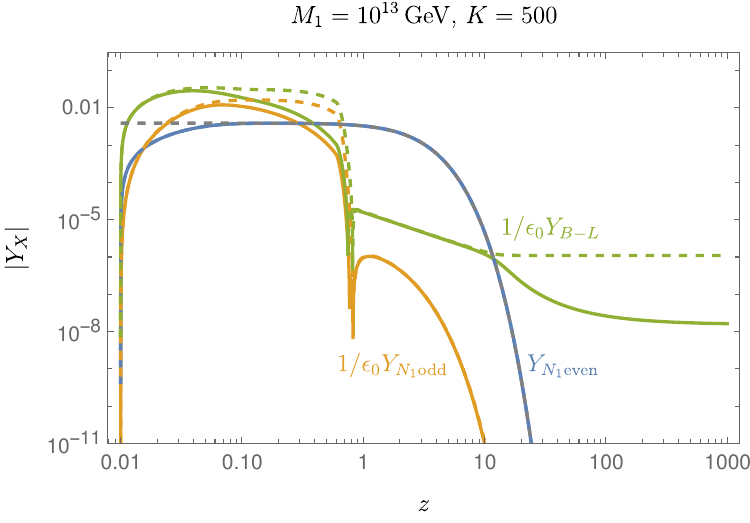}
\label{fig:washout_500}
}
\caption{Numerical solutions of Boltzmann equations with fully equilibrated spectators with (solid lines) and without (dashed lines) $\Delta L = 2$ processes for $\tilde{M}_1 = \SI{e13}{GeV}$ for $K=100$ (a) and $K=500$ (b).}
\label{fig:washout}
\end{figure}

\begin{figure}[t]
\centering
\subfloat[]{
\centering
\includegraphics[width=0.48\textwidth]{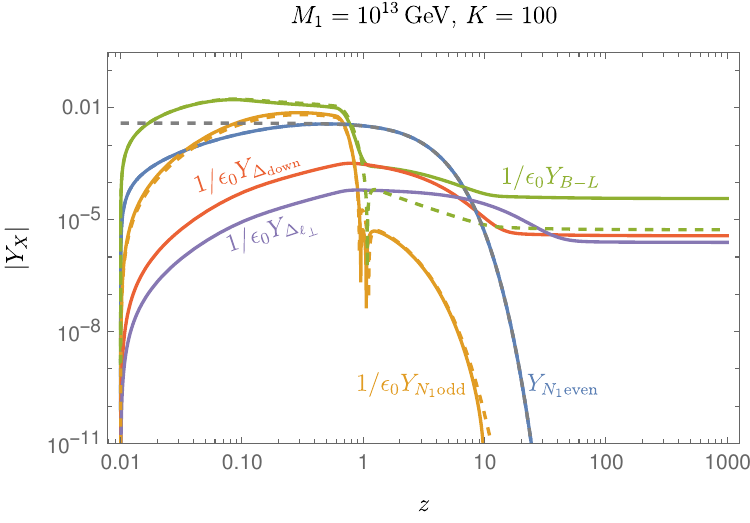}}
\subfloat[]{
\centering
\includegraphics[width=0.48\textwidth]{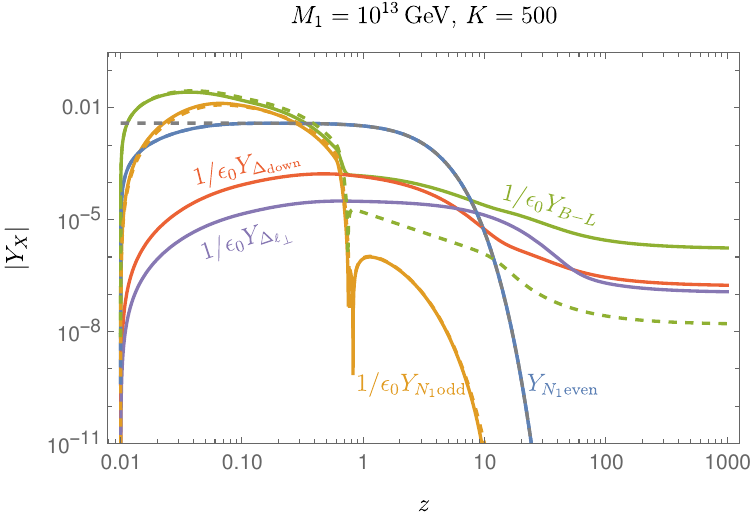}
\label{fig:spectators_500}}
\caption{Numerical solutions of Boltzmann equations with partially (solid lines) and fully equilibrated (dashed lines) spectators for $\tilde{M}_1 = \SI{e13}{GeV}$ for $K=100$ (a) and $K=500$ (b).}
\label{fig:spectators}
\end{figure}

In \cref{fig:rates}, we show the $\Delta L=2$ rates $W_2$ compared to the $\Delta L = 1$ rates $\eta_{N_1} \Gamma$. At large $z$, the rate $W_2$ scales as $z^{-2}$, in agreement with Refs.~\cite{Buchmuller:2002rq,Buchmuller:2004nz}, which is a slower suppression than the exponential Boltzmann suppression from $\eta_{N_1} \Gamma$. In this region, after all other processes have frozen out, the $\Delta L = 2$ processes continue to erase the resulting asymmetry. Numerical solutions of the Boltzmann equation with and without these new processes are shown in \cref{fig:washout}. In \cref{fig:spectators}, we present a comparison of the solutions with fully and partially equilibrated spectators in our new analysis including the $\Delta L=2$ rates. As can be seen from the plots, even with strong washout, the spectator processes substantially change the dynamics of the fields and can lead to differences of several orders of magnitude in the outcomes. In particular, the sign shift that happens at $z \approx 1$ with fully equilibrated spectators is absent once effects from partial equlibration of spectators and thereby the protection of early asymmetries from washout are included.

\begin{figure}[h!]
\centering
\subfloat[]{
\centering
\includegraphics[width=\textwidth]{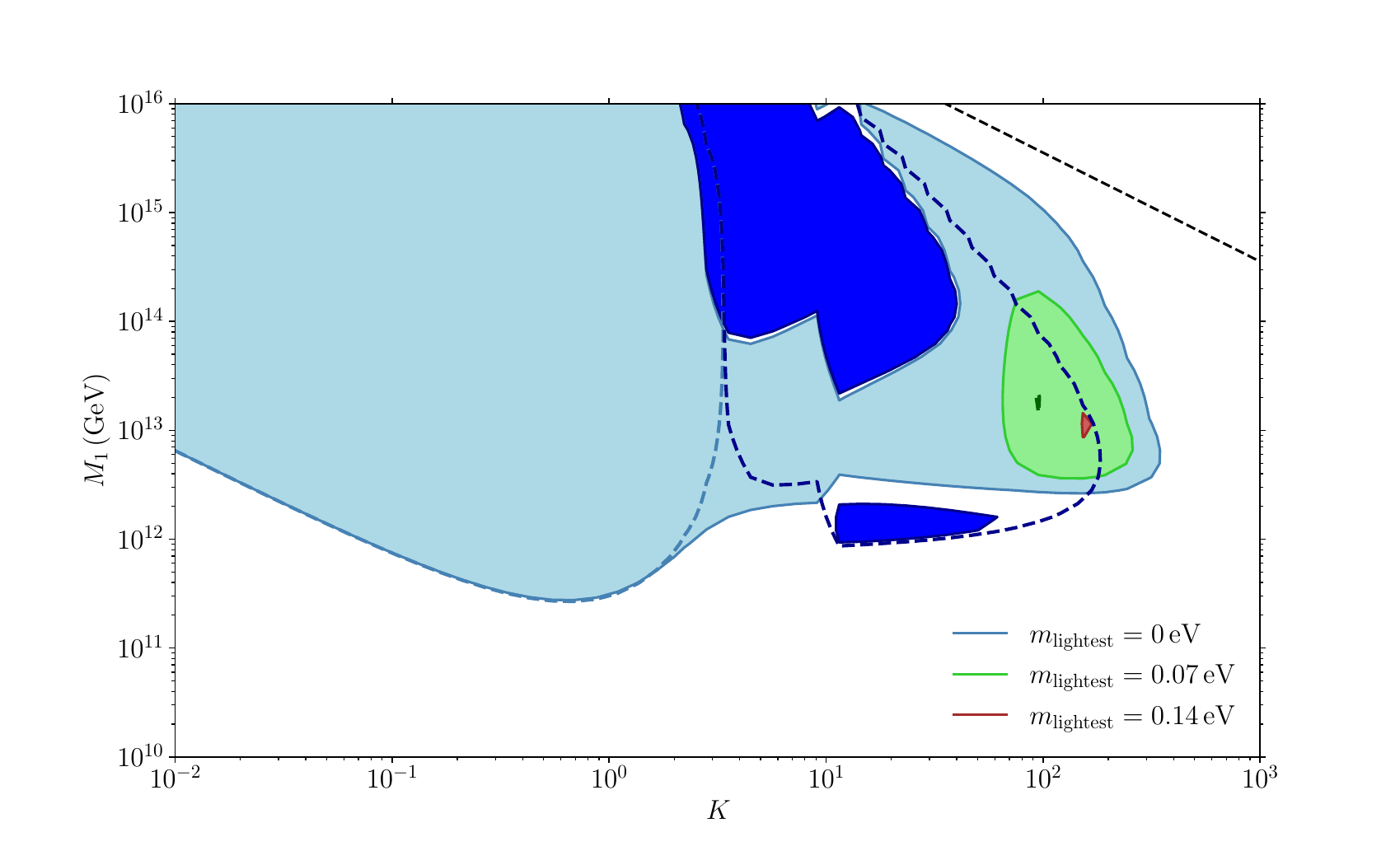}
\label{fig:vanishing}} \\
\subfloat[]{
\centering
\includegraphics[width=\textwidth]{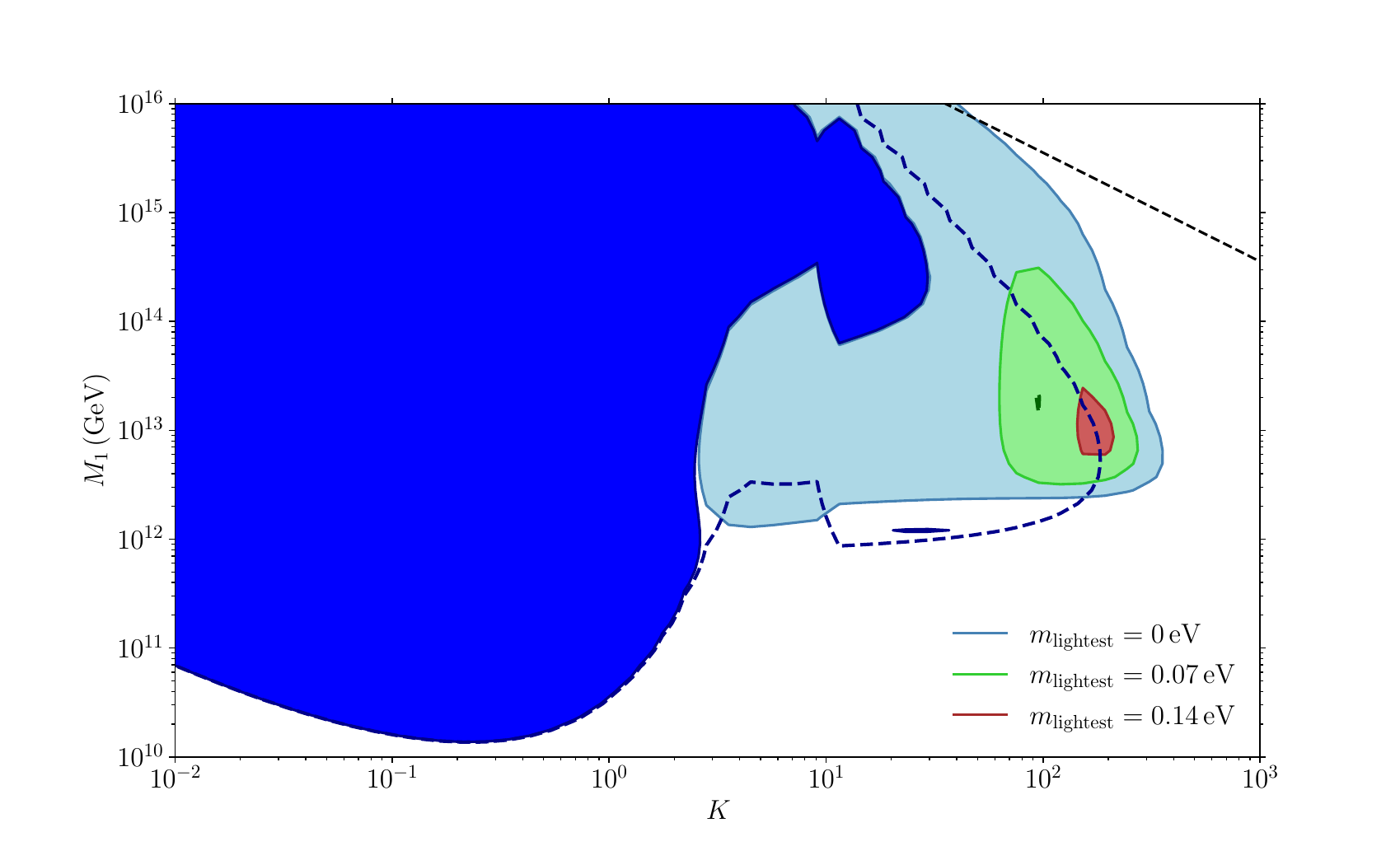}
\label{fig:thermal}}
\caption{Allowed regions for vanishing (a) and thermal (b) initial conditions and different choices of $m_\text{lightest}$. Lighter/darker shades represent different signs of the final asymmetry. The shaded regions are with partially equilibrated spectators, while the dashed contours are with fully equilibrated spectators.\label{fig:contours}}
\end{figure}

\section{Parameter Scan}
\label{sec:scan}

We solved the fluid equations numerically from $z=0.01$ to $z=1000$ for both vanishing and thermal initial conditions, varying $K$ between $10^{-2}$ and $10^3$ and $M_1$ between $\SI{e10}{GeV}$ and $\SI{e16}{GeV}$, and compare the scenarios with fully and partially equilibrated spectators. The results are shown in \cref{fig:contours}. With fully equilibrated spectators, we find that the previous bound of $m_\text{lightest} \lesssim \SI{0.12}{eV}$ gets tightened to $\SI{0.08}{eV}$, probably due to the improved treatment of the $\Delta L = 2$ processes leading to a larger rate than used in Ref.~\cite{Buchmuller:2003gz}. With partially equilibrated spectators, however, we find that the spectators protect part of the asymmetry from washout. This effect antagonizes the larger $\Delta L=2$ washout rates. Taking all this into account, sufficient asymmetries for $m_\text{lightest}$ as large as $\SI{0.15}{eV}$ can result. We further find a lower bound on $M_1$ of $\SI{e10}{GeV}$, which is slightly stronger than the Davidson-Ibarra bound~\cite{Davidson:2002qv}.

The impact of increasing the active neutrino masses is twofold. On the one hand, we have the constraint $K \propto \tilde{m}_1 \geq m_\text{lightest}$, which implies that increasing neutrino masses restricts $K$ to larger values, where washout is increasingly efficient. On the other hand, as discussed in Refs.~\cite{Buchmuller:2003gz,Hambye:2003rt}, increasing the neutrino masses also limits the asymmetry parameter. While one could circumvent this by increasing $M_1$, this would simultaneously increase the $\Delta L = 2$ washout rate, as pointed out in Refs.~\cite{Buchmuller:2002rq,Buchmuller:2003gz,Buchmuller:2004nz}, thereby also reducing the final asymmetry. With this, we do not find any allowed region for $m_\text{lightest} \geq \SI{0.15}{eV}$, which gives us an upper bound $m_\text{lightest} \lesssim \SI{0.15}{eV}$ for leptogenesis in this scenario.

The plots in \cref{fig:contours} are produced assuming normal hierarchy, but the difference to inverted hierarchy is negligible. The only term which is sensitive to the hierarchy is the maximal decay asymmetry. However, \cref{eq:asymmetry} is valid for both hierarchies, and the main difference between them is that $m_1$ and $m_3$ exchange roles as $m_\text{lightest}$. Doing the substitution $m_1 \leftrightarrow m_3$ leads to a relative minus sign in \cref{eq:asymmetry}, which can be compensated by a change in the sign of $y$. Therefore, the only practical difference between the two hierarchies is in the precise value of the absolute mass splitting between $m_1$ and $m_3$, which is negligible for our purposes.

\section{Conclusions}
\label{sec:conclusions}

In this work we have applied state-of-the-art techniques to unflavoured leptogenesis in a type-I seesaw model with hierarchical right-handed neutrinos. For this, we have derived $\Delta L = 2$ washout processes within the CTP-formalism and included them in our analysis, thus generalizing the computation from Ref.~\cite{Garbrecht:2019zaa}. We then used these methods to perform a parametric survey of a specific part of the parameter space for leptogenesis in the type-I seesaw mechanism. In particular we revisited the neutrino mass bound from Refs.~\cite{Buchmuller:2002jk,Buchmuller:2003gz,Giudice:2003jh}. The spectator fields significantly alter the evolution of the fields, leading to a relative sign change with respect to the scenario without spectators in large regions of the parameter space. They also protect the asymmetry from washout, leading to freeze-out asymmetries which are up to a few orders of magnitude larger than without them. With this, we obtain that the previously reported upper bound on the active neutrino masses for leptogenesis in this scenario is slightly relaxed, even though we use larger $\Delta L=2$ rates. While the change in the bound appears marginal because of the competing effects, the picture of the underlying dynamics is substantially altered by our use of the up-to-date methods.

As one reduces the mass of $N_1$, different spectator processes come into play, as well as flavour effects on the charged leptons. The same methods presented in this work could be applied to these scenarios, potentially leading to a sigificant change in the outcome of leptogenesis. In general, however, we expect spectator effects, in particular the interplay of their partial equilibration with early asymmetries, to enhance the freeze-out asymmetries, thus opening up viable parameter space for leptogenesis.

\section*{Acknowledgements}
We would like to thank Carlos Tamarit for very useful discussions and for sharing the code as well as the numerical results from Ref.~\cite{Garbrecht:2019zaa}. E.W.’s work is funded by the Deutsche Forschungsgemeinschaft (DFG, German Research Foundation) -- SFB 1258 -- 283604770. 

\appendix
\section{Review of the CTP formalism}

The principal idea of the CTP formalism is to perform a functional integration on a closed time contour, allowing one to compute and track expectation values of operators over time. With this we can, for instance, compute the evolution of particle number densities in the early Universe~\cite{Calzetta:1986cq}. As for the usual path-integral, we can compute $n$-point functions from the CTP path-integral, except that we now must distinguish between the branches going forward (''$+$'') and backward in time (''$-$''). The four CTP propagators for a complex scalar field $\phi$ are then~\cite{Prokopec:2003pj,Garbrecht:2018mrp,Beneke:2010wd,Beneke:2010dz}
\begin{subequations}
\begin{align}
	i \Delta_X^{++} (u, v) &= \braket{T \phi_X (u) \phi_X^\dagger (v)}, &  i \Delta_X^{--} (u, v) &= \braket{\bar{T} \phi_X (u) \phi_X^\dagger (v)}, \\
	i \Delta_X^< (u, v) &= \braket{\phi_X^\dagger (v) \phi_X (u)}, & i \Delta_X^> (u, v) &= \braket{\phi_X (u) \phi_X^\dagger (v)},
\end{align}
\end{subequations}
while for a fermion field $\psi$ we have, similarly,
\begin{subequations}
\begin{align}
	i S_{X, \alpha, \beta}^{++} (u, v) &= \braket{T \psi_{X, \alpha} (u) \bar{\psi}_{X, \beta} (v)}, &  i S_{X, \alpha, \beta}^{--} (u, v) &= \braket{\bar{T} \psi_{X, \alpha} (u) \bar{\psi}_{X, \beta} (v)}, \\
	i S_{X, \alpha, \beta}^< (u, v) &= -\braket{\bar{\psi}_{X, \beta} (v) \psi_{X, \alpha} (u)}, & i S_{X, \alpha, \beta}^> (u, v) &= \braket{\psi_{X, \alpha} (u) \bar{\psi}_{X, \beta} (v)}.
\end{align}
\end{subequations}

For an arbitrary two-point function $G^{ab}$ we define retarded and advanced two-point functions
\begin{subequations}
\begin{align}
	G^a = G^T - G^> = G^< - G^{\bar{T}} && G^r = G^T - G^< = G^> - G^{\bar{T}}
\end{align}
as well as spectral and Hermitian functions
\begin{align}
	G^\mathcal{A} = \frac{1}{2 i} (G^a - G^r) = \frac{i}{2} (G^> - G^<) && G^\mathcal{H} = \frac{1}{2} (G^a + G^r) = \frac{1}{2} (G^T - G^{\bar{T}}).
\end{align}
\end{subequations}

It is further useful to work in Wigner space, where the Wigner transform of a two-point function is defined as
\begin{equation}
	G(x, k) = \int d^4 r e^{i k r} G \left(x + \frac{r}{2}, x - \frac{r}{2} \right),
\end{equation}
which depends both on the momentum $k$ and on the center of mass coordinate $x$. The two-point functions satisfy Schwinger-Dyson equations~\cite{Calzetta:1986cq,Cornwall:1974vz}, which, in Wigner space, are given by~\cite{Prokopec:2003pj}
\begin{subequations}
\begin{align}
	e^{- i \diamond} \{k^2 - m_X^2 - \Pi_X^{a, r} \} \{\Delta_X^{a, r}\} &= 1,\\
	e^{- i \diamond} \{ k^2 - m_X^2 - \Pi_X^ r \}\{\Delta_X^{<,>}\} &= e^{- i \diamond} \{\Pi_X^{<,>}\}\{\Delta_X^a\}, \\
	e^{- i \diamond} \{ \slashed{k} - m_X - \Sigma_X^{a, r} \}\{i S_X^{a, r}\} &= i P_X, \\
	e^{- i \diamond} \{ \slashed{k} - m_X - \Sigma_X^ r \}\{i S_X^{<,>}\} &= e^{- i \diamond} \{\Sigma_X^{<,>}\}\{i S_X^a\},
\end{align}
\end{subequations}
with the diamond operator $\diamond$ defined as
\begin{equation}
	\diamond \{A(x, k)\} \{B(x,k)\} = \frac{1}{2} [(\partial_x^\mu A(x, k))(\partial_{k, \mu} B(x, k)) - (\partial_k^\mu A(x, k)) (\partial_{x, \mu} B (x, k))],
\end{equation}
and the scalar and fermionic self-energies
\begin{subequations}
\begin{align}
	\Pi_X^{ab} (u, v) =& i ab \frac{\delta \Gamma^{\text{2PI}}}{\delta \Delta^{ba} (v, u)}, \\
	\Sigma_X^{ab} (u, v) =& - i ab \frac{\delta \Gamma^{\text{2PI}}}{\delta S_X^{ba} (v, u)},
\end{align}
\end{subequations}
where the functional $\Gamma^{\text{2PI}}$ is minus $i$ times the sum of the two-particle irreducible vacuum graphs.

To zeroth order in the gradients and to leading order in the Yukawa couplings, the Schwinger-Dyson equations are
\begin{subequations}
\begin{align}
	\label{eq:SD-1}
	(k^2 - m_X^2 - \Pi_X^{a, r}) \Delta_X^{a, r} &= 1,\\
	\label{eq:SD-2}
	(k^2 - m_X^2 - \Pi_X^ r) \Delta_X^{<,>} &= \Pi_X^{<,>} \Delta_X^a, \\
	\label{eq:SD-3}
	(\slashed{k} - m_X - \Sigma_X^{a, r}) i S_X^{a, r} &= i P_X, \\
	\label{eq:SD-4}
	(\slashed{k} - m_X - \Sigma_X^r) i S_X^{<,>} &= \Sigma_X^{<,>} i S_X^a.
\end{align}
\end{subequations}
In kinetic equilibrium, \cref{eq:SD-2,eq:SD-4} give
\begin{subequations}
\begin{align}
	i \Delta_X^< (k) =& 2 \Delta_X^\mathcal{A} f_X (k), & i \Delta_X^> (k) =& 2 \Delta_X^\mathcal{A} (1 + f_X (k)), \\
	i S_X^< (k) =& - 2 S_X^\mathcal{A} (k) f_X (k), & i S_X^> (k) =& 2 S_X^\mathcal{A} (k) (1 - f_X (k)),
\end{align}
\end{subequations}
where $f_X (k)$ are equilibrium distributions with chemical potential $\mu_X$:
\begin{align}
	f_X (k) = \frac{1}{e^{\beta (k_0 - \mu_X)} + 1} \text{(fermions)}, && f_X (k) = \frac{1}{e^{\beta (k_0 - \mu_X)} - 1} \text{(bosons)}.
\end{align}
The spectral functions can be obtained from \cref{eq:SD-1,eq:SD-3}. At tree level, they are given by
\begin{subequations}
\begin{align}
	\Delta_X^{\mathcal{A}, \text{tree}} (k) =& \pi \delta(k^2 - m_X^2) \text{sign} (k_0), \\
	S_X^{\mathcal{A}, \text{tree}} (k) =& \pi \delta(k^2 - m_X^2) \text{sign} (k_0) P_X (\slashed{k} + m_X),
\end{align}
\end{subequations}
while the spectral functions with the loop insertions summed up are given by
\begin{align}
S_X^{\mathcal{A}} (k) = P_X \left[\left(\slashed{k} + m_X - {\Sigma}^\mathcal{H}_X \left(k \right) \right) \cdot \frac{\Gamma_X}{\Omega_X^2 + \Gamma_X^2}
- {\Sigma}^\mathcal{A}_X \left(k \right) \frac{\Omega_X}{\Omega_X^2 + \Gamma_X^2} \right], && \Delta_X^\mathcal{A} (k) = \frac{\Gamma_X}{\Omega_X^2 + \Gamma_X^2},
\end{align}
with 
\begin{align}
\Gamma_X \left(k\right) = 2 \left( k_{\mu} - \Sigma^\mathcal{H}_{X, \mu} \right) \cdot \Sigma^{\mathcal{A},\mu}_X, && \Omega_X \left(k\right) = \left( k_{\mu} - \Sigma^\mathcal{H}_{X, \mu} \right)^2 - m_X^2 - \left(\Sigma^{\mathcal{A}}_{X,\mu}\right)^2,
\end{align} 
for fermions and
\begin{align}
	\Gamma_X (k) = \Pi_X^\mathcal{A}, && \Omega_X (k) = k^2 - m_X^2 - \Pi_X^\mathcal{H},
\end{align}
for scalars.

A short comment on the derivation of \cref{eq:collision_expanded} is in order here. In order to go from \cref{eq:collision_term} to \cref{eq:collision_expanded}, we made use of the KMS relation
\begin{equation}
	\hat{\Sigma}_{N_1, L/R}^> (k) = - e^{(k_0 \mp \mu_{\ell} \mp \mu_\phi)/T} \hat{\Sigma}_{N_1, L/R}^< (k),
\end{equation}
which, inserted into \cref{eq:SD-2}, implies
\begin{equation}
	S_{N_1, L/R}^> (k) = - e^{(k_0 \mp \mu_{\ell} \mp \mu_\phi)/T} S_{N_1, L/R}^< (k).
	\label{eq:KMS_N}
\end{equation}
Taken at face value, this would mean that $N_1, \ell$ and $\phi$ are in chemical equilibrium, with $\mu_{N_1, L/R} = \pm \mu_{\ell} \pm \mu_\phi$, which would indeed be the case if $\ell$ and $\phi$ were in chemical equilibrium with $N_1$. As it turns out, however, $N_1$ is not in chemical equilibrium, and its out-of-equilibrium decays drive $\ell$ and $\phi$ out of equilibrium. The use of equilibrium distribution functions for $\ell$ and $\phi$ here is merely an approximation in order to obtain the off-shell piece of the propagator for $N_1$. We therefore emphasize that applying the KMS relation \cref{eq:KMS_N} to the on-shell part of the propagator does not make sense. However, concerning the off-shell part of the propagator, we can interpret this relation to apply to $\Sigma_{N_1}$ and to the fields therein, and not to $N_1$ explicitly. Since we do approximate the fields contained in $\Sigma_{N_1}$ as being in kinetic equilibrium, it does make sense to use \cref{eq:KMS_N} for this off-shell part. This is the origin of the additional factor of two, which is characteristic of $\Delta L = 2$ processes.

\bibliography{references}

\end{document}